%% file: paper.tex
\documentclass[conference,dvipsnames]{IEEEtran}
\IEEEoverridecommandlockouts
\usepackage{todonotes}
\usepackage{hyperref}
\usepackage[font=footnotesize,labelfont=bf]{caption}
\usepackage{subcaption}
\usepackage{array,multirow,graphicx}
\usepackage{lipsum}
\usepackage{mathtools}
\usepackage{xcolor}
\usepackage{algorithm}
\usepackage{algpseudocode}
\usepackage{cite}
\usepackage{amsmath,amssymb,amsfonts}
\usepackage{graphicx}
\usepackage{textcomp}
\usepackage{mdframed}
\usepackage[none]{hyphenat}
\def\BibTeX{{\rm B\kern-.05em{\sc i\kern-.025em b}\kern-.08em
    T\kern-.1667em\lower.7ex\hbox{E}\kern-.125emX}}

\hypersetup{colorlinks=true,allcolors=blue}

\newcommand\dcp{\text{dCP}}
\newcommand\cp{\text{CP}}

\begin{document}

\title{Application-Level Differential Checkpointing \\ for HPC Applications with Dynamic Datasets
\thanks{Part of the research presented here has received funding from
    the European Union’s Seventh Framework Programme (FP7/2007-2013) and
    the Horizon 2020 (H2020) funding framework under grant agreement no.
    H2020-FETHPC-754304 (DEEP-EST). The present publication reflects
    only the authors' views. The European Commission is not liable for
    any use that might be made of the information contained therein.}
}

\author{\IEEEauthorblockN{Kai Keller and Leonardo Bautista Gomez \\
\textit{Barcelona Supercomputing Center (BSC-CNS)}\\
Barcelona, Spain \\
\{kai.keller, leonardo.bautista\}@bsc.es}
}

\maketitle

\begin{abstract}
    High-performance computing (HPC) requires resilience techniques such as
    checkpointing in order to tolerate failures in supercomputers. As the
    number of nodes and memory in supercomputers keeps on increasing, the size
    of checkpoint data also increases dramatically, sometimes causing an I/O
    bottleneck. Differential checkpointing (\dcp{}) aims to minimize the
    checkpointing overhead by only writing data differences. This is typically
    implemented at the memory page level, sometimes complemented with hashing
    algorithms. However, such a technique is unable to cope with dynamic-size
    datasets.  In this work, we present a novel \dcp{} implementation with a
    new file format that allows fragmentation of protected datasets in order to
    support dynamic sizes. We identify dirty data blocks using hash algorithms.
    In order to evaluate the \dcp{} performance, we ported the HPC applications
    xPic, LULESH 2.0 and Heat2D and analyze them regarding their potential of
    reducing I/O with \dcp{} and how this data reduction influences the
    checkpoint performance. In our experiments, we achieve reductions of up to
    62\% of the checkpoint time.

\end{abstract}


\input{1-introduction}

\input{2-terminology}

\input{3-relatedWork}

\input{4-implementation}

\input{5-hashalgorithms}

\input{6-threshold}
\input{7-analysis}

\input{8-discussion}

\input{9-conclusion}

\section{Acknowledgements}

This project has received funding from the European Union’s Seventh Framework
Programme (FP7/2007-2013) and the Horizon 2020 (H2020) funding framework under
grant agreement no.  H2020-FETHPC-754304 (DEEP-EST); and from the European
Union’s Horizon 2020 research and innovation programme under the LEGaTO Project
(legato- project.eu), grant agreement No 780681.  The present publication
reflects only the authors' views. The European Commission is not liable for any
use that might be made of the information contained therein.


\bibliographystyle{IEEEtran}
\bibliography{biblio}

%

\end{document}

%% file: 1-introduction.tex
\section{Introduction}
\label{sec:intro}

High-performance computing (HPC) is a major tool for scientific research and
fast industrial development. Supercomputers have observed an exponential
increase in size and performance over the last couple of decades. Exascale
computing (i.e., $10^{18}$ floating point operations per second) is the next
frontier and it promises to bring orders of magnitude more computing power into
the hands of scientists. However, the exponential increase in computational
power also comes with a certain number of challenges; for instance, power
consumption and resilience are among the most pressing issues that need to be
addressed to reach such extreme computing scales. Indeed, the increasing number
of components in large-scale systems makes the machine more prone to failures,
reducing the mean time between failures (MTBF). At the same time, the amount of
data used in large HPC simulations is increasing exponentially. Failures in
supercomputers are usually handled through checkpoint and restart, by storing
the state of the computation in reliable storage, so that the application can
restart from the last saved state upon a failure. Unfortunately, the reduction
in the MTBF forces users to checkpoint at a higher frequency to reduce the
amount of re-computation to be done in case of failure. Simultaneously, the
checkpoint takes more time as the amount of data to save increases. This leads
to a steep reduction in system efficiency. In order to maintain high
productivity in supercomputers and large data centers, it is important to
reduce as much as possible the amount of data to be checkpointed to reliable
storage.

Differential checkpointing has been proposed in order to avoid re-writing
checkpoint data that is identical between two consecutive checkpoints (i.e., no
change of data). Previous research works have attempted to implement such a
technique by tracking dirty memory pages in the system and only updating those
within the checkpoint (\cp{}) files. While this method works, it is not always
efficient as many applications do re-write the exact same content (e.g., zero)
into the same memory cells. From the OS perspective, these memory pages have
changed as they are dirty, but in reality the content has not changed. Hashing
the memory pages to detect real changes has also been proposed. Unfortunately,
this technique also fails to detect unmodified datasets with dynamic sizes
(e.g., particles moving between domains) or datasets relocated in memory.


In this paper we have implemented a hash-based strategy in which we partition
the application datasets (not the memory pages) in blocks and keep track of the
changes of each block by comparing the corresponding hashes.  In addition, we
introduce a new file format that is capable of recognizing changes in data
blocks and simultaneously dynamically adapt to changes in the size of the
protected structure. We evaluate the collision robustness of multiple hash
algorithms and show that MD5 and CRC32 are viable solutions for differential
checkpointing. We integrate our implementation into the multilevel
checkpointing library FTI and evaluate it with three HPC applications. In our
measurements, we obtain up to 62\% reduction in checkpointing time in
comparison to traditional checkpointing. Furthermore, we propose a theoretical
model that predicts performance gains that could be obtained with our \dcp{}
technique.

The rest of this paper is organized as follows. Section~\ref{sec:terminology}
introduces the terminology of this paper. Section~\ref{sec:related} discusses
related work. Section~\ref{sec:implementation} introduces our implementation of
\dcp{}. Section~\ref{sec:hashing} explores the robustness of different hashing
algorithms. Section~\ref{sec:threshold} presents our analytical model to
predict performance gains. The results of our large-scale evaluation are
presented in Section~\ref{sec:evaluation}.  Section~\ref{sec:discussion}
discusses the strong points and limitations of this proposed technique and
finally, Section~\ref{sec:conclusion} concludes this paper.

%% file: 2-terminology.tex
\section{Terminology}
\label{sec:terminology}

The term \emph{incremental checkpointing} is used in the literature to denote
two different processes. To avoid confusion we would like to clarify what we
refer to when we use the terms incremental and differential checkpointing.

\paragraph{Definition of Incremental Checkpointing} We refer to incremental
checkpointing as to be the \emph{incremental completion of a \cp{} file}.
This technique serves primarily to avoid overhead caused by oversaturated
network channels. It may be used within applications that provide datasets,
that define the current program state, at different times. Thus, instead of
writing the whole \cp{} data at once, it is incrementally written during
some period of time, which reduces the stress on the network.

\paragraph{Definition of Differential Checkpointing} We refer to differential
checkpointing as to be the \emph{differential update of a \cp{} file}.
That is, the data blocks in the previous \cp{} file that by the time of a
subsequent \cp{} differ to the corresponding data block of the current
application state, will be replaced by the up-to-date data block. The rest of
the blocks (i.e., those that did not change) will not be updated.

%% file: 3-relatedWork.tex
\section{Related Work}
\label{sec:related}

Differential updates of data-states, data-dependencies or workflows exist in
several disciplines of HPC.  Depending on the case, there are various methods
that allow detection and logging of differences in data-structures and
workflows. An interesting example for such a logging mechanism and the
differential update of \cp{} data in data streaming applications is well
described within the web documentation of Apache Flink~\cite{apacheFlink}. The
update mechanism is based on  a so-called \emph{log-structured-merge}. The data
storage is based on a key-value pair. In order to update the \cp{} files, the
first and fast in-memory storage layer collects the updated keys inside a
\emph{memtable}. After a certain amount of data has been accumulated inside the
memtable, the data is flushed to a stable storage.  The flushed memtables are
now called \emph{sorted-string-tables} (sstables). At a certain point, the
various sstables will be merged into one sstable. This is performed
asynchronously to the streaming application execution by a dedicated process.
The merge consolidates redundant keys from different sstables.

Another important work to mention is the differential (de-)serialization in
SOAP implemented by Nayef Abu-Ghazaleh et al.~\cite{SOAP:DDS,4032007,1323489}.
SOAP (simple object access protocol) is a messaging protocol adequate for
server communications. It can be used to negotiate between different
application layer protocols that encode the messages into the XML format (e.g.
HTTP, RPC or SMTP). SOAP is a promising candidate to negotiate between
independent transfer protocols in high-performance parallel and distributed
computing (HPCD) environments~\cite{SOAP}.  A bottleneck of the messaging
workflow is the (de-)serializing of messages.  Serializing refers to the
conversion of in-memory data to ASCII text messages encoded in the XML format
prior to the sending of a message and de-serialization refers to the reverse
process after a message has been received.  Nayef Abu-Ghazaleh et al. present a
mechanism that uses checksums in order to identify redundant information in
consecutive messages prior to the de-serialization.  The similarities may be
inside the message contents or within the encoding XML structure. Using this
information, it is possible to skip de-serialization of unchanged message
sequences.


Besides the two framework-specific examples from above and other specific
implementations, non-specific implementations of \dcp{}, e.g. as linkable
libraries, exist at kernel level (compare C. Wang et.  al~\cite{5695644} or R.
Gioiosa et.  al.~\cite{1559961}), i.e.  transparent for the application
developer and in form of compiler plugins that analyze C/R capabilities inside
the application during compile time (see G. Bronevetsky et. al.
~\cite{5160999}). However, kernel-level checkpointing is not always efficient
and not much exist for HPC applications at user-level.

The library \emph{libckpt}~\cite{pbkl:95:lib} can be operated almost
transparently (i.e. without modifying the application code), but, it also
provides API functions that enable the user to determine the \cp{} behavior.
The API permits to specify the \cp{} data (i.e.  explicitly exclude or include
certain memory regions) or the \cp{} location (i.e. where inside the
application flow) and location of the \cp{} files (i.e. path on the file
system).  In order to detect data updates between consecutive \cp{}s, libckpt
employs the UNIX page protection mechanism. The library has knowledge about the
process virtual address space. All memory pages that correspond to the process
address space are set to read only via a call to \texttt{mprotect} with the
\texttt{PROT\_READ | PROT\_EXEC} flags specified.  After this, every store
operation to one of the protected pages will rise a segmentation fault signal
(\texttt{SIGSEGV}). This signal is caught by libckpt and appropriately handled.
The address of the page is then marked dirty and will be written to disk during
the next \cp{}.  Every differential \cp{} (Plank denotes this as incremental
\cp{}~\cite{pbkl:95:lib}) represents an extra file on the file system. However,
the user may specify a parameter in order to restrict to a maximum number of
files. When this amount is reached, the files are being merged into one. This
strategy has two major drawbacks. First, by protecting the whole address space
of the application, one incorporates data that is not necessarily needed for a
successful restart. That means that one may expect \cp{} files to be much
larger than necessary, hence a higher checkpoint overhead.  Second, many
applications update continuously all the datasets, which does not imply that
the value after the update differs from the one before (e.g., zeros in a
domain). In this case, the page protection mechanism will not lead to a
significant reduction of data in the \dcp{} files missing the goal
of such a feature.


Kurt B. Ferreira et al.~\cite{10.1007/978-3-642-24449-0_31, kurtThesis}
developed the library \emph{libhashckpt} on top of libckpt. Before the dirty
memory pages are written to disk, the hashes of this pages are compared to the
hashes that were generated by the time of the former \cp{}. Only if the hashes
differ, the page will be incorporated in the \dcp{} update. The version of
libckpt that is provided at~\cite{libckptWeb} is restricted to 32-bit kernels
and thus cannot be used on almost any cluster/supercomputer. Also, we
were not able to acquire the library libhashckpt by any means in order to compare it
with our implementation. Libhashckpt is the closest work to our proposal;
nonetheless there are multiple differences. First, libhashckpt is based on
classic PFS-based checkpoint-restart and not implemented in a multilevel
checkpointing library with asynchronous checkpointing, which involves a number
of differences (see Section~\ref{sec:implementation}).  Second, libhashckpt
produces a file for every checkpoint update, having to deal with a high number
of files, which applies high stress on the metadata servers.  Third,
libhashckpt does not adapt well for applications with dynamic dataset sizes.
Indeed, when datasets change in size they might be moved to other memory
locations or force other datasets to shift in the memory space, this will look
like a completely different dataset from the memory page perspective but in
reality they are just the same datasets that have either been displaced or
changed in size. This will force a complete rewrite of the full checkpoint
data, missing once again the goal of differential checkpointing.

This paper addresses all those issues, making this proposal the only general
purpose multilevel checkpointing library that implements a version of
differential checkpointing that adapts to datasets with dynamic sizes through a
user-level interface and that scales for large HPC applications.


%% file: 4-implementation.tex
\section{\dcp{} Implementation in FTI}\label{sec:implementation}

In section~\ref{sec:related} we saw several examples of logging mechanisms that
may detect and track differences in application states. The mechanisms can be
divided into two categories: tracking dirty pages (i.e. pages that was accessed
by a store operation) and tracking actual changes of data by checksum
comparison. Apache Flink and libckpt apply the first category.  SOAP and
libhashckpt implement both strategies. Although the hash based strategy has an
advantage over the dirty page approach, applying hashes over memory pages still
has the disadvantage of lacking the application perspective, offering only a
\emph{black box} perspective over the data. Application-level interfaces
allow us to see datasets on their own, and detect real changes, reagrdless of
whether they move to another memory region.

%
%


FTI is an application-level checkpointing library, with an API that provides
flexibility and allows user to flag datasets that need to be protected.  In
addition, FTI is a multi-level \cp{} library that offers 4 levels of increasing
reliability and FTI implements a dedicated process that performs
post-processing work for the more reliable \cp{} levels asynchronously to the
application processes.  In our implementation, the virtual address space of the
datasets will be partitioned into blocks of size $b$. We create hashes of these
blocks and keep them in memory. The hashes are created from the dataset
representation in memory immediately after a successful \cp{} and \emph{before}
the application continues its normal execution so that the hashes in memory
belong to the state of the dataset that is stored in the \cp{} files. The
hashes are also applied before an asynchronous work (e.g., RS encoding) is
done. FTI also creates hashes in order to ensure \cp{} file consistency upon
restart, but these two types of hashes are unrelated.




We do not adopt the method of creating new files for every \dcp{} update.
Instead, we take advantage of the existing FTI \emph{head} feature
~\cite{bautista2011fti}, that is, we may assign work that is related to FTI
processes (e.g., RS-encoding, flushing \cp{} files from local storage to the
PFS) to a dedicated process that can operate asynchronously to the application
flow. In order to update the \cp{} file safely, we create a copy of the file
using the head process and update the copy. The former \cp{} file is kept
during the update in order to roll back to it when an error occurs during the
update.  After successful completion of the \cp{} the old \cp{} file will be
removed. This procedure prevents the corruption of the \cp{} data in the case
of a failure during the update process and minimizes the number of files on the
PFS which in turn reduces the stress on the metadata server.

\subsection{Dealing with Dynamic Sizes of Datasets}
\label{sec:fff}

In order to implement an efficient and scalable \dcp{} mechanism, the FTI
protected datasets need to be arranged in immutable blocks in the \cp{} files.
For protected datasets with steady sizes, this is accomplished naturally.
However, FTI supports datasets with dynamic sizes. Thus, to maintain immutable
positions inside the checkpoint file we need to allow fragmentation of the
datasets and in order to read the files upon the restart, we need to store the
dataset-file mapping.

\begin{figure}[htbp]
    \centerline{\includegraphics[width=\linewidth]{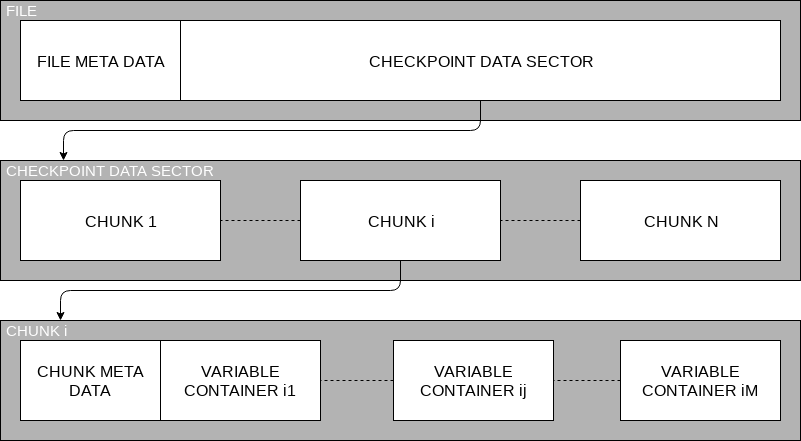}}
    \caption{FTI-FF structure: The \emph{File Meta Data} contains information
    that is used by FTI for other purposes than differential checkpointing
    (e.g., RS-encoding). The \emph{Chunk Meta Data} holds the file-mapping
    metadata for the dataset chunks stored in the current block (i.e.,
    chunk-size, container-size, id)}
\label{fig:ftiff-structure}
\end{figure}

In the current release, FTI stores metadata that is needed for the restart in
separate files. These files are getting parsed using the Iniparser
library~\cite{iniparser}. We could potentially extend this practice in order to
keep track of the data-file mapping. However, FTI writes one metadata file for
each group of processes (consult~\cite{bautista2011fti} or ~\cite{fti} for
details about the FTI library).  In order to minimize the overhead, we decided
to develop a file format for FTI, \emph{FTI-FF}, that includes the metadata for
the owning process within the file structure. The general file structure is
shown in figure~\ref{fig:ftiff-structure} (For a comprehensive description
please visit~\cite{FFF}).



\subsubsection*{FTI-FF Structure}

The file structure is generated dynamically. By the time of the first
successful \cp{}, every dataset has created a virtual container with the
current size of the dataset.  The container is located at an immutable position
in the \cp{} file next to the corresponding metadata block.  When a dataset
increases its size in one of the successive \cp{}s, hence exceeds the size of
the first virtual container belonging to this dataset, another virtual
container is created with the excess as its size. The new virtual container is
appended at the end of the file, tailing the corresponding metadata block.
This mechanism repeats every time a dataset exceeds the total size of the
existing virtual containers. Once created, a container never changes its
size. The total size of all the virtual containers belonging to the same
dataset will be seen as contiguous and will be filled linearly. This may
lead to sparse files when the data size shrinks.

\subsection{Updating the \cp{} Files}

The \emph{prime directive} we have to meet when implementing any \dcp{}
approach is that we must not update the data in the existing \cp{} files
directly.  This is due to the danger of corrupting the file if an error occurs
during the update.

In FTI we meet this goal by creating a copy of the \cp{} file after the
successful creation. The duplication is performed by a dedicated process
asynchronously to the application run so that the application processes can
continue the execution as if they would without \dcp{} functionality.  During
the next \cp{}, the processes may now update the copy directly. After the
successful completion, the former file can be deleted. If the dedicated process
cannot create the copy, the application processes will be notified and a
complete \cp{} will be created.

\subsection{Tracking the differences}

We mentioned earlier that the memory regions of the datasets will be
partitioned into blocks of size $b$ and that the content of the blocks is
represented by hashes. If two hashes that correspond to the same block differ,
we assume that the contents differ (\emph{dirty} blocks) and if the hashes
coincide, we assume that the contents are identical (\emph{clean} blocks).  We
have to distinguish between blocks that are old (\emph{valid}), i.e.  present
in the \cp{} file, and blocks that are new (\emph{invalid}), i.e. not present
in the \cp{} file (for instance, by the time of the first \cp{}, all blocks are
invalid). Invalid blocks will be added to the \cp{} file without hash
comparison.

In order to decide which data needs to be updated in the \cp{} files, we apply
the following set of rules:

\begin{itemize}
    \item [(I)] mark new blocks as invalid.
    \item [(II)] identify dirty blocks during the \dcp{} update.
    \item [(III)] update the \cp{} file with dirty or invalid blocks.
    \item [(IV)] crate/update hashes for invalid/dirty blocks.
\end{itemize}

\paragraph*{I}

When datasets are exposed to FTI, the corresponding blocks are marked invalid
to ensure that new datasets will be included in the \cp{} file.  The same
applies when datasets increase their size and new data blocks are exposed to
FTI. However, the hashes for the blocks will not be created yet.

\paragraph*{II}

During the \dcp{} update, the processes request contiguous dirty regions by
calling the function \texttt{FTI\_ReceiveDcpChunk()}. A dirty region is the
accumulation of adjacent dirty blocks.  The function takes a pointer to the
origin of the dataset and a size argument and compares sequentially the hashes
of blocks with size $b$ (user defined granularity). The function returns 1 and
updates the pointer with the base address of a dirty region and sets the size
of the region. 0 is returned if no dirty region was found. Invalid blocks
cannot be compared in that sense since they do not have a representation inside
the \cp{} file. Hence, invalid blocks will be included in dirty regions ad-hoc.

\paragraph*{III}

\texttt{FTI\_ReceiveDcpChunk()} is called inside a while loop and the
\cp{} copy is updated with the dirty regions returned by the function. The
loop continues until the function returns 0, signaling that the dataset
is now again up-to-date in the \cp{} file. At the first \cp{}, all blocks
will be written.

\paragraph*{IV}

After the successful completion of the \dcp{} update, the hashes that
correspond to dirty or invalid blocks will eventually be updated (or created
for invalid blocks) so that the hashes represent the actual state of the
datasets in the current \cp{} file\footnote{One could think it would be more
efficient to update the hash array during II, however, this would violate the
prime directive since we cannot assure that the \dcp{} update will be indeed
successful.}. Clearly,  we keep the hashes of blocks that are neither dirty nor
invalid untouched.

\begin{figure}[htbp]
    \centerline{\includegraphics[width=\linewidth]{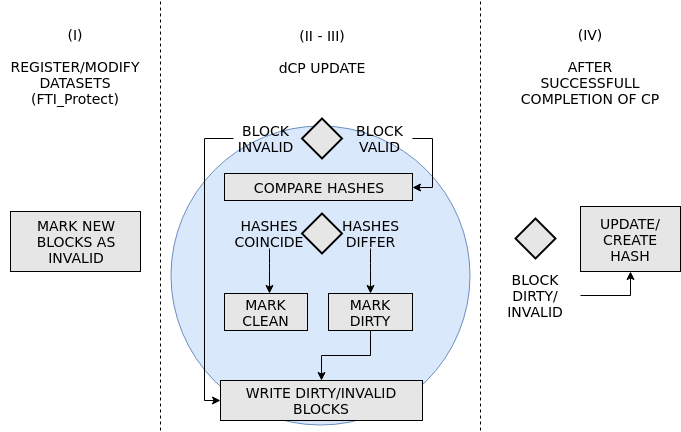}}
    \caption{\dcp{} detection and update scheme. Processes left to the
    blue circle happen before and processes to the right after the
    \dcp{} update. The circle indicates the dirty region request loop}
    \label{fig:checkpoint-circle}
\end{figure}

The process is visualized in figure~\ref{fig:checkpoint-circle}. The figure is
divided in three sections separated by a dashed line. The left section
corresponds to I and is implemented in function \texttt{FTI\_Protect}.  The
function is used in FTI to register datasets in order to include them into the
\cp{} files.  FTI creates metadata related to the dataset within this function.
After the first call to \texttt{FTI\_Protect}, all blocks of the corresponding
datasets are marked invalid. After a subsequent call in order to increase the
size of a dataset, blocks in the memory region that exceed the former size are
new to FTI and thus consequently marked invalid as well. This ensures that new
blocks are automatically included in the \cp{} files.  The middle section of
the figure corresponds to II and III and the third section corresponds to IV.
Hash creation (if invalid) or the update (if dirty), only happens in the third
section after the successful completion of the \dcp{} update (which corresponds
to a full \cp{} at the first call).

The relevant metadata for the \dcp{} mechanism is kept inside an array of the
structure \texttt{struct FTIT\_HashBlock}.  The array has $N$ elements, where
$N$ is the next greater integer of the dataset size divided by the block size
(or this very value if the dataset size is a multiple of the block size). Every
array element corresponds to one block of the partitioned dataset. The
structure has three members: A boolean that indicates if the hash is valid, a
boolean that indicates if the hash is dirty and the hash digest (either a
32-bit unsigned integer for CRC32 or a 128-bit unsigned char buffer for MD5).

%% file: 5-hashalgorithms.tex
\section{Choice of the Hash Algorithm}
\label{sec:hashing}

Depending on the size of the protected datasets, the hash arrays
might get significantly large. For instance, the MD5 digest
length is 128 bits (16 bytes). Assuming a hash-block size of 128 bytes
and 1GB of protected data per rank, we have to reserve 128MB of RAM for
the hash metadata. In order to reduce this size, we can either increase
the block size or decrease the digest size. The former may decrease
the \dcp{} performance due to the coarser resolution (i.e., more dirty
block updates) and the latter may increase the risk for inconsistent
\cp{} files due to higher collision rates of the hash algorithm (i.e.,
when a collision occurs the block is considered clean despite the fact
that the data in the block has changed).

In order to provide a small digest size, we tested three 32-bit hash
algorithms (digest size 32 bits) upon performance and reliability.
Adler32, Fletcher32 and CRC32. For completeness, we included also MD5
(digest size 128 bits) in the tests although it is considered to be
reliable and fast for data integrity checks (despite its flaws in the
cryptographic area~\cite{HARRAN2017}). The Adler32 and CRC32 checksums
were calculated using the zlib data compression library~\cite{zlib},
Fletcher32 was implemented using the recommendations
in~\cite{Nakassis1988} and for MD5 we used the OpenSSL
library~\cite{openssl}.

Fletcher32 and Adler32 are both
significantly faster than CRC32. However, both also have poor collision
resistant characteristics for block sizes that are relevant in our case,
as we will see below. To obtain a statement about the reliability of the
checksums we performed a simple collision test.  We focussed on the 
so-called \emph{avalanche effect}~\cite{avalanche}, since in real
applications it is very possible that elements of the datasets change
only very little.  The test follows algorithm~\ref{alg:collision}.

\begin{algorithm}
\caption{Count hash collisions of modified buffers}\label{alg:collision}
\begin{algorithmic}
\Repeat{
    \ForAll{ $b$ }
        \State populate $C_b$ with $N_b$ random u64 integers;
        \State create hashes $h_{C_b}$ of $C_b$;
        \ForAll{$p$}
            \For{i=1, $N_b$}
                \State $D_{b,i}$ = $C_{b,i}$ $\oplus$ $p$;
                \State Create hash $h_{D_{b,i}}$ of $D_{b,i}$;
                \If{ $h_{D_{b,i}}$ == $h_{C_{b,i}}$ }
                \State $c_{b,p}++$;\Comment{$c_{b,p}$ := Collision Counter}
                \EndIf
            \EndFor
        \EndFor
    \EndFor
}
\Until{N iterations}
\end{algorithmic}
\end{algorithm}

$C_b$ and $D_b$ are buffers that contain random integers, $b
= \{2^i \, | \, 7 \leq i\leq 15\}$ denotes the hash block sizes and $p$
denotes the patterns that are used to modify the elements of $C_b$.  For
$N_b$ we have $b \bmod (N_b\times 64) == 0$.  The elements of $p$
correspond to bit flips of the last 1 ($p_0=\,$\texttt{0x1}), 2
($p_1=\,$\texttt{0x3}), 4 ($p_2=\,$\texttt{0xff}), 8
($p_3=\,$\texttt{0xfff}) and 16 (\texttt{$p_4=\,$0xffff}) bits and to an
arbitrary modification ($p_5$ is an arbitrary pattern) to simulate a
random change.

Fletcher32 is commonly implemented with $M=2^n$ or $M=2^n-1$ ($M$ is the
modulo value for the checksum. Consider~\cite{Nakassis1988} for
implementation details). The case $M=2^n-1$ leads to identical checksums
for buffers that differ only in one or more groups of two consecutive
bytes that are all \texttt{0x00} in one and all \texttt{0xff} in the
other buffer. For us, this is reason enough to disqualify the algorithm
for its usage in \dcp{}.  Nevertheless, we included it in our
measurements.

\begin{table}[htb]
    \caption{Collision rates (i.e. the probability of collision per iteration)
    achieved by application of algorithm~\ref{alg:collision}. We did not detect
    any collision for CRC32 or MD5 and the collision rates for Fletcher32
    mod(65535) were almost identical to Fletcher mod(65536). Thus, we do not
    list the results here. For all cases, the number of iterations have been
    within 160-180 million.}
\label{tab:collisions}
\resizebox{\linewidth}{!}{\scriptsize
    \begin{tabular}{lllllll}
\hline\noalign{\smallskip}
 & $p_0$      & $p_1$    & $p_2$  & $p_3$ & $p_4$ & $p_5$ \\
\noalign{\smallskip}
\hline
\noalign{\smallskip}
$b$     & \multicolumn{6}{c}{ADLER32} \\
\noalign{\smallskip}
\hline
\noalign{\smallskip}
128  &6.84e-3&1.42e-3&8.56e-5& 3.68e-7&	6.13e-9&	1.23e-8 \\
256  &1.70e-3&3.46e-4&2.12e-5& 8.59e-8&	1.23e-8&	1.23e-8 \\
512  &4.24e-4&8.69e-5&5.39e-6& 1.84e-8&	0&6.13e-9 \\
1024 &1.06e-4&2.21e-5&5.39e-6& 1.84e-8&	0&	6.13e-9 \\
2048 &2.56e-5&5.21e-6&2.58e-7&0&	6.13e-9&	0 \\
4096 &6.23e-6&1.37e-6&9.20e-8&	0&	6.13e-9&0\\
8192 &1.56e-6&2.70e-7&1.84e-8&	6.13e-9&0&0\\
16384&3.56e-7&4.91e-8&4.29e-8&	6.13e-9&	0&0\\
32768&1.41e-7&7.98e-8&1.84e-8&	0&	0&	0\\
\noalign{\smallskip}
\hline\noalign{\smallskip}
        & \multicolumn{6}{c}{FLETCHER32 - MOD(65536)} \\
\noalign{\smallskip}
\hline
\noalign{\smallskip}
128   &1.54e-5	&1.47e-5	&1.52e-5	&1.52e-5	&1.50e-5	&1.54e-5\\
256   &1.53e-5	&1.55e-5	&1.53e-5	&1.54e-5	&1.56e-5	&1.54e-5\\
512   &1.48e-5	&1.56e-5	&1.53e-5	&1.52e-5	&1.53e-5	&1.52e-5\\
1024  &1.48e-5	&1.55e-5	&1.51e-5	&1.53e-5	&1.58e-5	&1.56e-5\\
2048  &1.49e-5	&1.51e-5	&1.49e-5	&1.49e-5	&1.50e-5	&1.56e-5\\
4096  &1.57e-5	&1.53e-5	&1.57e-5	&1.53e-5	&1.51e-5	&1.50e-5\\
8192  &1.55e-5	&1.51e-5	&1.49e-5	&1.54e-5	&1.47e-5	&1.54e-5\\
16384 &1.51e-5	&1.55e-5	&1.52e-5	&1.54e-5	&1.55e-5	&1.52e-5\\
32768 &1.56e-5	&1.59e-5	&1.48e-5	&1.57e-5	&1.53e-5	&1.52e-5\\
\hline
\noalign{\smallskip}
\end{tabular}}
\end{table}

The results of the collision test are listed in table~\ref{tab:collisions}.
Adler32 and Fletcher32 exhibit a significant amount of collisions. Most of the
collisions for Adler32 occurred for 1-bit or 2-bit flips and decrease for
increasing block sizes.  The collisions for Fletcher32 are homogeneously
distributed for all modifications and block sizes. We estimate the collision
rate of both algorithms, Adler32 and Fletcher32, as being too high in order to
provide a sufficient level of reliability. MD5 and CRC32, on the other hand, did
not show any collisions. The test we have performed is not appropriate to
deliver a solid cryptographic statement about the reliability of CRC32 and MD5,
however, it is enough to disqualify Adler32 and Fletcher32 for our purpose.
Based on literature about CRC32 and MD5 (CRC32 is used in zlib and other cases
to provide data integrity~\cite{GUERON2012179,zlib,zlibWeb}) and based on our
results we are quite confident about its application for \dcp{}.

%% file: 6-threshold.tex
\section{When is Differential Checkpointing Beneficial?}
\label{sec:threshold}

In order to estimate the threshold at which differential checkpointing becomes
beneficial, we construct a cost function from the reduction in CP overhead:

\begin{equation}\label{eq:saving-th}
    \Delta T_{s} =\lvert N_d t_w - N_t t_w \rvert\,  = (N_t-N_{d})\,t_{w} ,
\end{equation}

and from the additional generated overhead (i.e. the time to determine the differences):

\begin{equation}\label{eq:hash_differences}
    \Delta T_{o} =(N_{t}+N_{d})\,t_{h} .
\end{equation}

$t_w$ is the duration to write a block of data with block-size $b$,
$t_h$ the duration of hashing the block, $N_d$ is the number of blocks
that differ and $N_t$ is the total number of blocks. The saving in equation~\ref{eq:saving-th} corresponds to the absolute value of the
time difference between writing all blocks ($N_t t_w$) and writing
only the dirty blocks ($N_d t_w$). The overhead in
equation~\ref{eq:hash_differences} corresponds to the time to hash
the data blocks.
Equation~\ref{eq:hash_differences} involves both values, $N_t$ and
$N_d$, since, we cannot commit the new hashes
for data-blocks that differ prior to the successful completion of the
\cp{}, hence we compute these twice\footnote{We may avoid the
redundancy here if we store the hashes for the \textit{dirty} blocks in
a separate array, which would lead to a higher memory footprint.}.
After normalizing to the total number of blocks $N_t$ we get:

\begin{equation}\label{eq:threshold-bare}
    \tau  = (t_h-t_w)+n_d\, (t_w+t_h) , \quad n_d=N_d/N_t.
\end{equation}

Where $\tau := \Delta T/N_t = (\Delta T_o - \Delta T_s)/N_t$.
Equation~\ref{eq:threshold-bare} can be considered a cost
function that turns into a reduced overhead (speedup) for $\tau < 0$ and to additional overhead for $\tau > 0$. We can infer, that the maximal overhead
accounts to $2N_t t_h$ when $n_d = 1$. This corresponds to a maximal relative
overhead of $2t_h/t_w$ (i.e., relative to the time without \dcp{}). 

We may define the threshold, $\eta$,  at $\tau=0$ as:

\begin{equation}\label{eq:threshold}
    \eta := n_d \biggr|_{b, \tau = 0} = \frac{t_w-t_h}{t_w+t_h}  \approx \frac{1-\rho}{1+\rho} , \quad \rho = \frac{t_h}{t_w}.
\end{equation}

Let us keep in mind that $\tau$ depends on the block size $b$ as well.
$\eta$ corresponds to the threshold ratio of updated \cp{}
data (i.e. dirty) to the total amount of \cp{} data below which we
can expect a speedup. Equation~\ref{eq:threshold} is defined for $\eta
\in [0,1]$ and behaves monotonic in that regime. The lower the value for
$\rho$ the closer $\eta$ gets to 1, which would correspond to a
threshold of $n_d=1\mathrel{\widehat{=}}$ 100\% dirty (i.e. no overhead). 

We can give an estimation of $\eta$ by comparing the time that it takes
to write and that it takes to hash a block of data. Thus, we
measure the time, $t_w$, to write a block of size $b$ to disk, where we
consider the write to be a collective operation. We do so by measuring
the total time, $T_w$, to write a buffer of size $n*b$ and computing
$t_w=T_w/n$. $T_w$ is the time for a collective write (i.e. all
processes must have finished I/O). And also we measure the time, $t_h$,
that it takes to compute the hash for a block of size $b$.  In contrast
to $t_w$, $t_h$ is computed by $t_h=\sum t_{h,i}/n$, where $t_{h,i}$ is
the time to hash the ith block, thus $t_h$ is the average value of all
$t_{h,i}$. This reflects in contrast to $T_w$ a local
(non-collective) operation.

Since the hash creation is local to the ranks we may expect a perfect scaling
behavior for $t_h$. For $t_w$ instead, we have to consider network bottlenecks
that can slow down the I/O processes towards a larger scale. Thus,
we expect an increasing speedup for increasing total problem sizes.

\begin{figure}
\includegraphics[width=\linewidth]{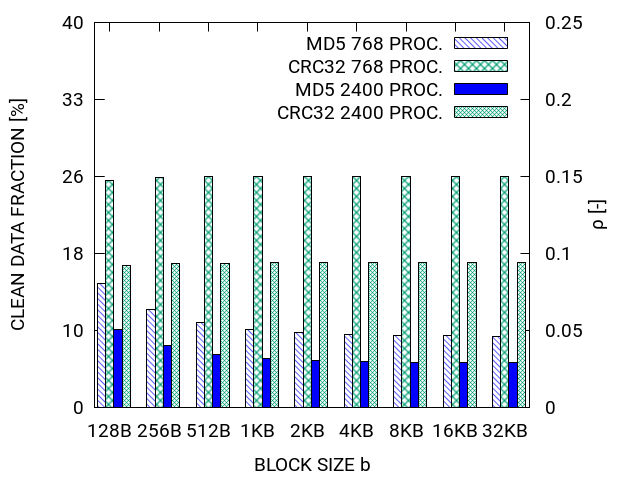}
\caption{The bars show the estimated \dcp{} threshold, i.e. the fraction
    of clean data we need to make the \dcp{} operation beneficial. The
    left axis shows the clean data fraction (1-$\eta$), the right axis
    shows the value of $\rho$ (ratio between the hash time, $t_h$, and
    I/O time, $t_w$, for block size $b$) that corresponds to the respective
    value of $1-\eta$ on the left axis.  The experiment has been
    performed with 768 and 2400 processes and 1GB per rank.}
\label{fig:threshold-lower-bound}
\end{figure}

Figure~\ref{fig:threshold-lower-bound} shows the results for the measurements
we performed for 768 and 2400 processes. In both cases, the total buffer size
was 1GB per process which leads to the total problem sizes of 0.75 TB and 2.3
TB respectively.

Note that $\eta$ is the threshold w.r.t the dirty blocks.
In figure~\ref{fig:threshold-lower-bound} we show $1-\eta$, which corresponds to the threshold w.r.t the clean blocks. We can see that the threshold indeed decreases for a growing problem size. We observe a better
performance of MD5 towards CRC32 in all cases. The performance of MD5
depends slightly on the hash-block size. This dependency is less strong
at a larger scale. This also applies for the performance difference
between CRC32 and MD5. The results show that for $b=32$KB and MD5, the threshold is at merely 5\% clean data share (i.e. 5\% less to write).

%% file: 7-analysis.tex
\section{Evaluation}
\label{sec:evaluation}

In section~\ref{sec:threshold} we have seen that even when applications update
95\% of the checkpoint data (i.e. we save only about 5\% of I/O) \dcp{} can
already be beneficial for HPC applications. In order to demonstrate this
theoretical result with empirical evidence, we analyze the behavior of \dcp{}
in FTI while checkpointing three HPC applications at large scale. We conduct
representative experiments that analyze performance and overhead.  All
experiments were performed on MareNostrum4.  Each node is composed
by~\cite{mn4userguide,mn4press}: \\

\begin{itemize}
    \item 2 Intel Xeon Platinum 8160 CPU (24 cores at 2.10GHz)
    \item 12 $\times$ 8 GB DDR4-2667 DIMMS (96GB/node)
    \item 100 Gbit/s Intel Omni-Path HFI Silicon 100 Series PCI-E
    \item 10Gbit Ethernet
    \item 200 GB SSD local to the nodes
    \item SUSE Linux Enterprise Server 12 SP2
\end{itemize}

\subsection{HPC Applications}
\label{sec:apps}

In this section we introduce the different applications used during our large
scale evaluation.

\subsubsection{LULESH 2.0}
\label{par:lulesh}

Lulesh~\cite{LULESH2:changes} is part of the Advanced Simulation and Computing
(ASC) program from the Lawrence Livermore National Laboratory (LLNL). It
simulates a Sedov blast wave propagation within one material in three
dimensions~\cite{LULESH:spec}.  The modeling space is discretized into an
unstructured hex mesh. The system state is updated using stencil operations.
The purpose of LULESH is to provide a proxy application that possesses the
characteristics of an HPC application from this field in order to analyze
performance on various platforms and various programming models. That makes it
very interesting for us as an example since it represents a broad field of
applications.




In order to maximize the checkpoint load, we conducted measurements that
determined the highest value that we can pass to LULESH without the risk of a
memory overflow on the node. The checkpoint data is serialized, which increases
the memory footprint of the application. With a \cp{} size of 430MB per rank,
we use about 80GB of the node memory (96GB available) and achieve an aggregate
\cp{} size of 725GB.

\subsubsection{xPic}
\label{par:xpic}

xPic is an alternative implementation of the physical problem treated in
iPic3D~\cite{MARKIDIS20101509}. iPic3D and xPic are part of the application
co-design in the DEEP-EST project~\cite{DEEPEST}. The
application models space plasma simulations. The modeling space is discretized
by a rigid mesh. The mesh is defined in the configuration file. The simulation
is always initialized to the equilibrium state. In each time step, the particle
states and electromagnetic fields are advanced using the Vlasov equation, which
couples the equation of motion to the Maxwell equations.

xPic takes its runtime parameters from a configuration file. In order to scale
the problem size, we used a combination of the parameters \texttt{ntcx} (number
of cells in x-direction), \texttt{ntcy} (number of cells in y-direction) and
\texttt{nppc} (number of particles per cell). To control the number of
contiguous datasets, we modified the parameters \texttt{nblockx} (number of
blocks in x-direction), \texttt{nblocky} (number of blocks in y-direction) and
\texttt{nspec} (number of species).  We implemented two distinct mechanisms in
order to expose datasets to FTI. In the first implementation, xPic-c (c for
common), we expose every memory contiguous dataset individually to FTI.
Depending on the configuration of xPic, this may lead to a large number of
protected variables. In the second implementation, xPic-s (s for serialized),
we use BOOSTs \texttt{libboost\_serialization} library~\cite{bib:boost} to
combine the datasets into one contiguous buffer which is then exposed to
FTI.

\subsubsection{Heat2D}
\label{par:heat2d}

Heat2D is a 2D heat distribution simulation using a 1D domain decomposition.
It simulates the transition from a non-equilibrium heat distribution to the
equilibrium state. In each time step, the cells of the temperature grid are
updated via a 4-point stencil operation that stores the average of the 4
neighbor cells temperatures into the center cell. The ranks exchange adjacent
rows of the temperature grids. The simulation runs until the total value of the
temperature differences reaches a pre-defined minimal value or exceeds a
certain number of iterations.  The large majority of memory used by Heat2D is
checkpointed which enabled us to perform large scale executions with large
checkpoitn sizes, for instance a run with a total problem size of about 2.8TB
with 2304 processes on 48 nodes.

\subsection{Variation of the Block Size $b$}\label{sec:blocksize}

We start by analyzing the impact of the block size over the effectiveness of
\dcp{}. We measured the time of a \dcp{} update for various block sizes $b$
and compared the results to an ordinary \cp{} (\dcp{} disabled).  All \cp{}s
were performed at the same application state. We performed experiments with
both MD5 and CRC32, the results were very similar for both hashing algorithms
thus we show only the MD5 results for space constrains.  By decreasing the
block size, we increase the granularity. That means that we have a better
chance to get close to the actual percentage of data that did change. This
should result in fewer data to write and therefore we expect better performance
with smaller blocks.

\begin{table}[htb]
\centering
\caption{Impact of the block size $b$ on the \dcp{} update time for xPic using MD5. Negative
    values of $\tau$ correspond to a speedup and positive values to overhead.
    \emph{HASH SIZE} lists the respective memory sizes that the hash
    tables occupy in memory. The problem size was 1568MB per rank.}
\label{tab:diff-b-sizes}
\resizebox{\linewidth}{!}{\scriptsize
\begin{tabular}{lp{1.0cm}p{1.0cm}p{1.0cm}p{1.0cm}p{1.0cm}}
\hline\noalign{\smallskip}
    $b$ & $\tau$ & \dcp{} RATE & SHARE HASH & SHARE WRITE & HASH SIZE~[MB] \\
    \noalign{\smallskip}\hline\noalign{\smallskip}
    128B & \textcolor{Red}{1333\%} & 52.25\% & 1.51\%   & 97.67\% & 196   \\
    256B & \textcolor{Red}{1106\%} & 53.84\% & 1.53\%   & 97.39\% & 98    \\
    512B & \textcolor{Red}{666\%} & 56.25\%  & 2.10\%   & 96.13\% & 49    \\
    1KB &  \textcolor{Red}{231\%} & 59.15\%   & 4.40\%   & 91.40\% & 25    \\
    2KB &  \textcolor{Red}{15\%} & 61.42\%    & 12.82\%  & 73.93\% & 12    \\
    4KB &  \textcolor{ForestGreen}{-32\%} & 62.25\%     & 21.77\%  & 55.07\% & 6    \\
    8KB &  \textcolor{ForestGreen}{-35\%} & 62.41\%     & 22.69\%  & 52.47\% & 3    \\
    16KB & \textcolor{ForestGreen}{-36\%} & 62.48\%    & 22.66\%  & 52.52\% & 1.5  \\
    32KB & \textcolor{ForestGreen}{-36\%} & 62.50\%    & 22.67\%  & 52.07\% & 0.76 \\
    \noalign{\smallskip}\hline
\end{tabular}}
\end{table}


Table \ref{tab:diff-b-sizes} shows the results for the experiment we performed
with the xPic application (see \ref{par:xpic} for details).  The first column
of the table shows the block size and the third column shows the percentage of
data written compared to the original checkpoint size.  We notice that as the
block size increases, the amount of data to write increases as well, due to the
lower granularity.  However, the overhead (shown in the second column) is
incredibly large for high block granularities (i.e., small blocks). To
understand this phenomena, we measured the time spent hashing and the time
spent writing data for each case.  We observe that the large majority of
checkpointing time is spent in writing and not hashing. This is caused by the
fragmentation of the updates into small chunks. It has been shown in the past
(e.g.  ~\cite{4536277,Shan2007UsingIT,NvmeWopsss2018}), that PFSs have poor
performance when small chunk sizes need to be written.



For xPic, block sizes of less than 4KB degrade performance and block sizes
greater than 4KB improve performance up to 36\%.  In addition, we measured the
amount of memory consumed to store the hash tables. Most of the block sizes
have hash tables that represent less than 1\% of the memory used by the
process. For block sizes of 16KB the hash tables take only 0.1\% of the memory
used by the application. Based on this analysis, we decided to use block sizes
of 16KB during the following measurements.


\subsection{Spatial and Temporal Differences}
\label{sec:spatiotemporal}

After finding the right block size to avoid too coarse hashes as well as
to fine I/O writes, we investigate the amount of data that is actually
being updated between two consecutive checkpoints for the applications
presented in Section~\ref{sec:apps}.

\begin{figure*}[tbh]
    \centering
    \includegraphics[width=.32\textwidth]{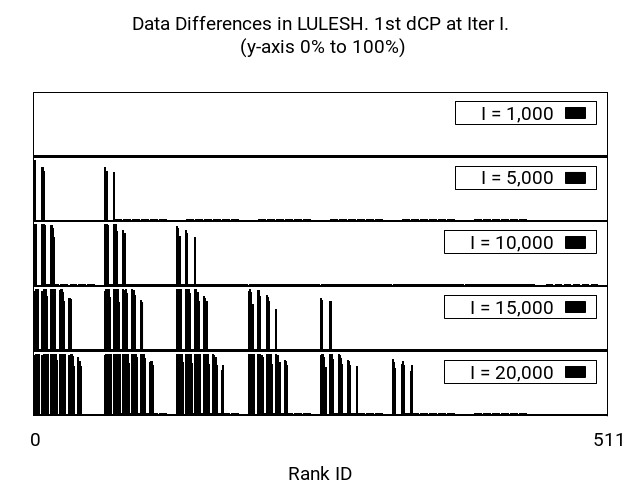}
    \includegraphics[width=.32\textwidth]{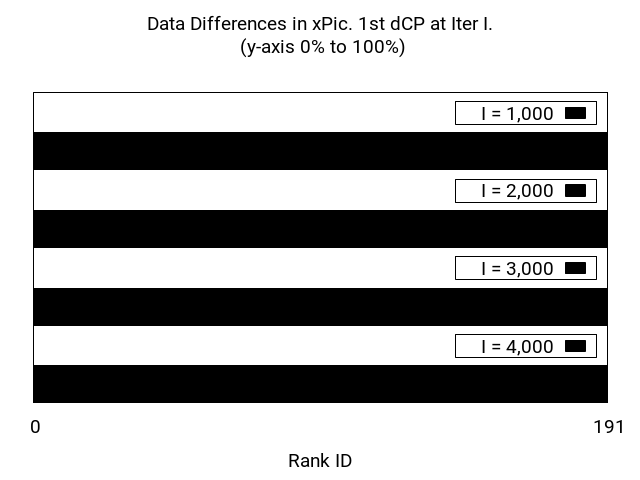}
    \includegraphics[width=.32\textwidth]{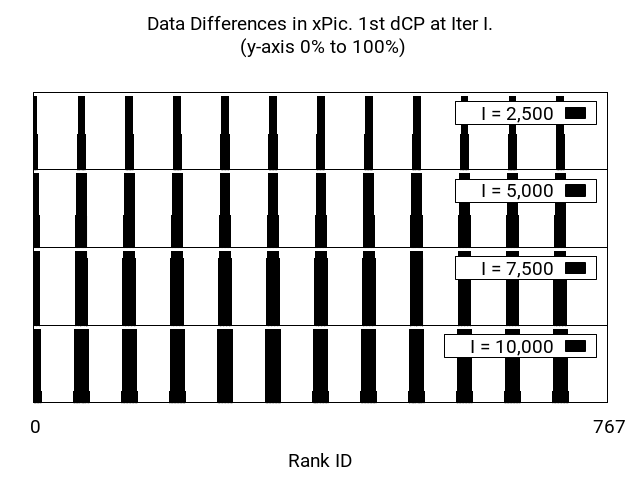}
    \caption{Data differences after first checkpoint for different ranks (x
    axis) and different times for checkpointing (y axis).}
    \label{fig:ckptDiff}
\end{figure*}

The results are depicted in Figure~\ref{fig:ckptDiff}. The three sub-figures are
divided into several temporal regions following the y axis (i.e., \dcp{} taken
at iteration $1000$, $5000$, etc.) and spatial regions following the x axis
(i.e., the process rank which is representative of a slice of the domain).
First, we observe that LULESH does not change too much data during the first
iterations; and as the time passes (up to iteration $20000$) the number of ranks
where data is actually modified increases. This reflects the shock wave that is
simulated by LULESH. This demonstrates that for applications like LULESH, the
benefits of \dcp{} might vary depending on time and space.

xPic on the other hand, shows a completely different behavior, the amount of
data updated is consistently the same across all the ranks and regardless of
the time in the execution. This is explained by the fact that xPic is a plasma
simulation in which particles are constantly in movement, even in those
changes are minimal, they are enough to trigger updates as they will produce a
different block hash. There are a few variables of the application that are
read-only and that do not change through out the simulation, which is why not a
100\% of the data is updated at every checkpoint.

Looking into Heat2D, we observe a middle ground between LULESH and xPic.
Indeed, Heat2D also increases the data differences as time evolves, but a much
lower pace than LULESH, giving it a less dynamic look. We observe that the most
affected ranks are organized in strides, which is consistent with the 1D
partitioning mentioned in Section~\ref{par:heat2d}. However, other initial
conditions could translate into a more homogeneous updates across ranks.

\subsection{Overhead reduction on HPC Applications}

In this section we evaluate the overhead of \dcp{} in comparison with
classic \cp{} for the three applications.

Table~\ref{tab:lulesh-times} lists the results of our measurements performed
with LULESH. The first row represents the full \cp{} and the second a
checkpoint with \dcp{} in which only 3\% of the data is updated. We have only
two rows since we never had updates significantly different to 3\%.  This
result indicates that the propagation of the wave is slow as shown previosly.
This great reduction in checkpoint size with \dcp{} in LULESH translates into a
62\% reduction in the \cp{} time.


\begin{table}[htb]
\centering
    \caption{Relative overhead of \dcp{} compared with full \cp{} for LULESH.
    Negative values correspond to a reduction of the overhead (speedup) and
    positive values to an increase in the overhead.}
    \label{tab:lulesh-times}
    \resizebox{\linewidth}{!}{\scriptsize
    \begin{tabular}{p{1.8cm}p{1.8cm}p{1.8cm}p{1.8cm}}
        \hline
        \noalign{\smallskip}
        \multicolumn{4}{c}{Relative checkpoint overhead compared to full \cp{} ($\Delta T / T_0$ [\%] )} \\ 
        \noalign{\smallskip}
        \hline\noalign{\smallskip}
        Data diff. ($n_d$)   & MD5            & CRC32         & NO \dcp{} \\
        \noalign{\smallskip}
        \hline
        \noalign{\smallskip}
        100\%       &\textcolor{ForestGreen}{-9$\pm$12}  & \textcolor{ForestGreen}{-5$\,\pm\,$13} & \textcolor{ForestGreen}{-5$\,\pm\,$13} \\
        3\%      & \textcolor{ForestGreen}{-62$\,\pm\,$10}      & \textcolor{ForestGreen}{-60$\,\pm\,$8}     & -              \\
        \hline
    \end{tabular}
    }
\end{table}




For xPic, we evaluate the non-serialized as well as the serialized
implementations (xPic-c and xPic-s, see~\ref{par:xpic}) were each
implementation is tested against two distinct configurations (A and B).  For
configuration A, the FTI protected memory consist of many relatively small
contiguous datasets.  Configuration B instead has few but rather large
contiguous datasets.  Table~\ref{tab:xpic-config} summarizes the relevant
runtime parameters for both configurations.

\begin{table}[htb]
    \centering
    \caption{Dataset sizes for the various xPic configurations.}
    \label{tab:xpic-config}
    \resizebox{\linewidth}{!}{
    \begin{tabular}{lllll}
        \hline\noalign{\smallskip}
        & \multicolumn{2}{c}{CONFIG. A} & \multicolumn{2}{c}{CONFIG. B} \\
        \noalign{\smallskip}\hline\noalign{\smallskip}
        & xPic-c           & xPic-s           & xPic-c           & xPic-s           \\
        \noalign{\smallskip}\hline\noalign{\smallskip}
        SIZE OF DATASETS [MB]     & 4.22           & 1360           & 168              &  1344.25         \\
        \# OF DATASETS \hspace{0.5cm} & 320 & 1 & 8  &  1   \\
        CP SIZE / RANK [MB]       & 1350.32        & 1360.38        & 1344.55          &  1344.80         \\
        CP SIZE TOTAL [GB]          & 760            & 765            & 882              &  883        \\
        \noalign{\smallskip}\hline
    \end{tabular}
    }
\end{table}

The results of our evaluation with these configurations is shown in
Table~\ref{tab:xpic-times}. First, we observe that the reduction on checkpoint
size is the same for executions with and without  serialization.  Another
observation is that the application of \dcp{} for configuration A does not
reduce the checkpoint overhead. The reason for this is that configuration A
produces a large number of small chunks to be written.  A more detailed
analysis of this phenomena is done in section \ref{sec:discussion}.  In
contrast, we do observe an important overhead reduction for configuration B.
The best performance measured is for xPic-s (serialized) using MD5 with up to
35\% speedup while writing only 50\% of the original checkpoint size.

\begin{table}[H]
    \centering
    \caption{Relative overhead of \dcp{} compared with full \cp{} for xPic.
    Negative values correspond to a reduction of the overhead (speedup) and
    positive values to an increase in the overhead.}
    \label{tab:xpic-times}
    \resizebox{\linewidth}{!}{\scriptsize
    \begin{tabular}{p{0.9cm} p{1.5cm} p{0.8cm} p{0.8cm} p{0.9cm}}
    \hline\noalign{\smallskip}
    \multicolumn{5}{c}{Relative checkpoint overhead compared to full \cp{} ($\Delta T / T_0$ [\%] )} \\ 
    \noalign{\smallskip}
    \hline\noalign{\smallskip}
    &    Data diff. ($n_d$) & MD5             & CRC32           & NO \dcp{}  \\
    \noalign{\smallskip}
    \hline
    \noalign{\smallskip}
    \parbox[t]{5mm}{\multirow{2}{*}{xPic-c (A)}}
        &100\%        & \textcolor{Red}{7$\,\pm\,$11}      & \textcolor{Red}{6$\,\pm\,$12}      & 0$\,\pm\,$9  \\
        & 50\%       & \textcolor{Red}{9$\,\pm\,$12}      & \textcolor{Red}{11$\,\pm\,$9}   & - \\
    \noalign{\smallskip}
    \hline
    \noalign{\smallskip}
    \parbox[t]{5mm}{\multirow{2}{*}{xPic-c (B)}}
        &100\% & \textcolor{Red}{9$\,\pm\,$16} & \textcolor{Red}{14$\,\pm\,$11} & \textcolor{ForestGreen}{-3$\,\pm\,$9}     \\
        & 62\%       & \textcolor{ForestGreen}{-33$\,\pm\,$6}      & \textcolor{ForestGreen}{-28$\,\pm\,$6}   & -  \\
    \noalign{\smallskip}
    \hline
    \hline
    \noalign{\smallskip}
    \parbox[t]{5mm}{\multirow{2}{*}{xPic-s (A)}}    &100\%        & \textcolor{Red}{7$\,\pm\,$17}      & \textcolor{Red}{14$\,\pm\,$9}      & 0$\,\pm\,$7  \\
    & 50\%       & \textcolor{ForestGreen}{-4$\,\pm\,$6}      & 0$\,\pm\,$6   & - \\
    \noalign{\smallskip}
    \hline
    \noalign{\smallskip}
    \parbox[t]{5mm}{\multirow{2}{*}{xPic-s (B)}} &100\% & \textcolor{Red}{5$\,\pm\,$5} & \textcolor{Red}{11$\,\pm\,$7} & \textcolor{ForestGreen}{-2$\,\pm\,$6}     \\
    & 62\%       & \textcolor{ForestGreen}{-35$\,\pm\,$7}      & \textcolor{ForestGreen}{-29$\,\pm\,$6}   & -  \\
    \noalign{\smallskip}\hline
    \end{tabular}}
\end{table}

As mentioned in Section~\ref{sec:spatiotemporal}, the data difference in Heat2D
depend significantly on the initial conditions. Heat2D shows a good reduction
of checkpoint size, in the regime of 40\% to 100\%.  Table \ref{tab:heat2d}
lists the results. We can see that MD5 has clearly performance benefits in
comparison to CRC32. We notice that almost all of the experiments show an
significant reduction on the checkpoint overhead. We observe important speedups
of up to 49\% for a 40\% \dcp{} update using MD5.

\begin{table}[htb]
    \centering
    \caption{Relative overhead of \dcp{} compared with full \cp{} for Heat2D.
    Negative values correspond to a reduction of the overhead (speedup) and
    positive values to an increase in the overhead.}
    \label{tab:heat2d}
    \resizebox{\linewidth}{!}{\scriptsize
    \begin{tabular}{p{1.5cm}p{1.5cm}p{1.5cm}p{1.5cm}}
    \hline\noalign{\smallskip}
    \multicolumn{4}{c}{Relative checkpoint overhead compared to full \cp{} ($\Delta T / T_0$ [\%] )} \\ 
    \noalign{\smallskip}
    \hline\noalign{\smallskip}
    Data diff. ($n_d$)   & MD5            & CRC32         & NO \dcp{}   \\
    \noalign{\smallskip}
    \hline
    \noalign{\smallskip}
100\%   & \textcolor{ForestGreen}{-2$\,\pm\,$9}   	& \textcolor{Red}{1$\,\pm\, $6}   & \textcolor{ForestGreen}{-4$\,\pm\,$11}  \\
99\%	& \textcolor{ForestGreen}{-5$\,\pm\, $7}  	& \textcolor{ForestGreen}{-2$\,\pm\, $7}  	& -		\\
95\%	& \textcolor{ForestGreen}{-8$\,\pm\, $6}  	& \textcolor{ForestGreen}{-7$\,\pm\, $7}  	& -		\\
87\%	& \textcolor{ForestGreen}{-14$\,\pm\, $6}      & \textcolor{ForestGreen}{-12$\,\pm\, $6}  	& -		\\
79\%	& \textcolor{ForestGreen}{-19$\,\pm\,$8} 	    & \textcolor{ForestGreen}{-17$\,\pm\, $6}  	& -		\\
71\%	& \textcolor{ForestGreen}{-26$\,\pm\, $6}  	& \textcolor{ForestGreen}{-22$\,\pm\, $6}  	& -		\\
63\%	& \textcolor{ForestGreen}{-35$\,\pm\, $5}   	& \textcolor{ForestGreen}{-30$\,\pm\, $5}   & - 	\\
56\%	& \textcolor{ForestGreen}{-40$\,\pm\,  $5}  	& \textcolor{ForestGreen}{-37$\,\pm\, $4}  	& -		\\
40\%    & \textcolor{ForestGreen}{-49$\,\pm\,  $5}    	& \textcolor{ForestGreen}{-46$\,\pm\, $7} 	& - 	\\
    \hline
    \end{tabular}
    }
\end{table}

Overall, the three applications (although with different behaviours) show
substantial improvements thanks to \dcp{}. The reduction in checkpointing
overhead goes up to 62\%, 35\% and 49\% for LULESH, xPic and Heat2D
respectively.

%% file: 8-discussion.tex
\section{Discussion}\label{sec:discussion}

In section~\ref{sec:threshold} we presented a theoretical model that
may be used to estimate the speedup we may achieve using \dcp{}. In this
section, we want to check whether the predictions from the model
coincide with the measurements or not.

Let us write down the relative time difference, $S$,  of a \dcp{} update towards
a conventional \cp{}:
\[
    S(n_d) = \Delta T(n_d)/T_0 :=\,
        \begin{cases}
            < 0 : \text{overhead reduction} \\
            > 0 : \text{overhead increase} \\
        \end{cases}.
\]
$T_0$ denotes the time for a full \cp{} and \dcp{} disabled.
Using equation~\ref{eq:threshold-bare} we may write this as:
\begin{equation}\label{eq:tau-model}
    S(n_d) = \frac{\tau}{t_w} = \rho-1+n_d (\rho +1) \quad , \quad \rho = \frac{t_h}{t_w}
\end{equation}
We used here that $T_0 = t_w N_t$. We determined $t_w$ and $t_h$ by a
separate measurement and the values we use here are:
\begin{align}
    b &= 16\text{KB} \\
    t_w &= 1.35\times10^{-3}s \\
    t_h &= 3.92\times10^{-5}s \quad \left[\text{MD5}\right] \\
    \rightarrow \rho &= 0.029
\end{align}
For clarity, we will consider only the results for MD5.

\begin{figure}[htb]
\centering
    \includegraphics[width=\linewidth]{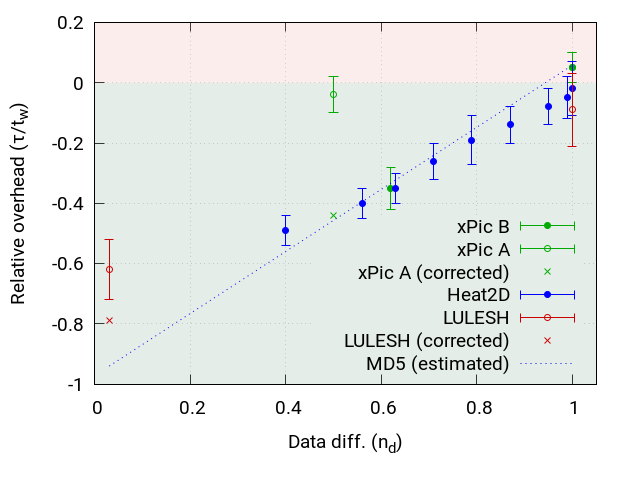}
    \caption{Measured and estimated speedup/overhead of \dcp{} updates. The
    green background indicates the region where we have speedup and the red
    region indicate overhead. $\tau/t_w=0$ corresponds to the threshold (i.e.,
    the full \cp{} baseline) The datasets with the label \emph{corrected},
    refer to measurements that used a buffer to collect small chunks in order
    to avoid small chunck writes.}
    \label{fig:tau-model}
\end{figure}

Figure~\ref{fig:tau-model} shows the measured relative speedups and the
estimation computed by equation~\ref{eq:tau-model}. The figure shows that in
two cases the estimation is near to accurate and in two cases it is not as
accurate.  Heat2D and xPic with configuration B show both a very good matching
to the estimation done with the theoretical model. LULESH shows the highest
speedup with 62\%, however, using equation~\ref{eq:tau-model} we would expect a
speedup of about 94\%. For xPic-s with configuration A, we measured a 4\%
speedup but expected about 46\%.

\begin{figure}[htb]
\centering
    \includegraphics[width=\linewidth]{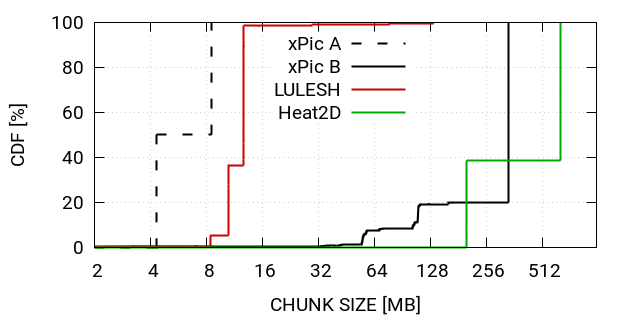}
    \caption{Cumulative distribution function (CDF) for chunk sizes of
    contiguous dirty regions during \dcp{} updates.}
    \label{fig:tau-model-frac}
\end{figure}

Given the disagreement between theoretical prediction and experimental results
for LULESH and xPic with configuration A, we performed a more detailed
analysis.  Figure~\ref{fig:tau-model-frac} shows the cumulative density
function (CDF) of chunk sizes written contiguously during a \dcp{} update for
all four scenarios. The figure reveals a correlation between the size of the
chunks and the performance. xPic-s A and LULESH show both less performance than
expected and both write mostly chunks of relatively small sizes (4MB - 12MB).
On the other hand, we have good performance in xPic B and Heat2D where we
observe relatively large chunk sizes (mostly over 200MB).

If the small writes are the explanation for the inaccurate model predictions,
one should be able to meet the estimated performance by avoiding I/O operations
with small chunks. This can be accomplished by allocating a buffer of
sufficient size and collecting small chunks in this buffer until the
accumulated chunk size exceeds an appropriate I/O size (e.g., 16MB), to then
finally write to a dedicated file. We implement this technique and redo the
experiments.

The results for LULESH and xPic after implementing this modification are
denoted as \emph{corrected} in figure~\ref{fig:tau-model-frac}.  We can see
that after correction xPic A is in very good agreement with the model
prediction. LULESH also improved but it is still not as good as the model has
predicted. We continue the detailed analysis of LULESH and we noticed that
LULESH has about 2-3\% updates in all ranks \emph{except} in rank 0. Rank 0 has
a \dcp{} share of 80\%. A large amount of the data is thus written by only one
rank. It appears, that the model performs less well if the distribution of the
\dcp{} share is highly anti-symmetric. For all the other cases, we have a very
good matching between model prediction and experimental results, if we avoid
small chunk sizes in I/O operations.

%% file: 9-conclusion.tex
\section{Conclusion}
\label{sec:conclusion}

In this paper we experimented with the UNIX page protection mechanism in order
to determine data differences, but this mechanism is not able to differ between
assignments that leave the data invariant and assignments that indeed change
the data, since every access to a memory address always causes this address to
be considered as \emph{dirty} by the operating system.  Thus, we implemented a
hash-based differential checkpointing mechanism capable to detect real data
changes. We tested 4 hash algorithms upon performance and reliability and our
conclusion is that CRC32 and MD5 are safe choices to implement \dcp{}.

Another challenge of implementing differential checkpointing is that some
applications have datasets that change in size during the execution.  To
overcome this issue, we developed a new file format (FTI-FF) for FTI that
includes its own metadata and is extensible.  Our results indicate that the
\dcp{} performance, for our prototype implementation, is significantly better
in most of the cases. For some applications, it might depend on the chunk size
of contiguous dirty blocks that are written to the \cp{} files during the
\dcp{} update. Indeed, we observed a better performance towards larger chunk
sizes, and \dcp{} becomes clearly inefficient for very small chunk sizes.
However, we have demonstrated that this issue can be resolved by a mechanism
that collects small chunks into a large block until an appropriate accumulated
size is reached before writing to stable storage.  We observed a speedup of up
to 49\% in Heat2D, 35\% in xPic and 62\% in LULESH.

%

%% file: paper.bbl
\begin{thebibliography}{10}
\providecommand{\url}[1]{#1}
\csname url@samestyle\endcsname
\providecommand{\newblock}{\relax}
\providecommand{\bibinfo}[2]{#2}
\providecommand{\BIBentrySTDinterwordspacing}{\spaceskip=0pt\relax}
\providecommand{\BIBentryALTinterwordstretchfactor}{4}
\providecommand{\BIBentryALTinterwordspacing}{\spaceskip=\fontdimen2\font plus
\BIBentryALTinterwordstretchfactor\fontdimen3\font minus
  \fontdimen4\font\relax}
\providecommand{\BIBforeignlanguage}[2]{{%
\expandafter\ifx\csname l@#1\endcsname\relax
\typeout{** WARNING: IEEEtran.bst: No hyphenation pattern has been}%
\typeout{** loaded for the language `#1'. Using the pattern for}%
\typeout{** the default language instead.}%
\else
\language=\csname l@#1\endcsname
\fi
#2}}
\providecommand{\BIBdecl}{\relax}
\BIBdecl

\bibitem{apacheFlink}
``Managing large state in apache flink: An intro to incremental
  checkpointing,''
  \url{https://flink.apache.org/features/2018/01/30/incremental-checkpointing.html},
  accessed: 2018-04-23.

\bibitem{SOAP:DDS}
N.~Abu-Ghazaleh and M.~J. Lewis, ``Differential checkpointing for reducing
  memory requirements in optimized soap deserialization,'' in \emph{The 6th
  IEEE/ACM International Workshop on Grid Computing, 2005.}, Nov 2005, pp. 6
  pp.--.

\bibitem{4032007}
------, ``Lightweight checkpointing for faster soap deserialization,'' in
  \emph{2006 IEEE International Conference on Web Services (ICWS'06)}, Sept
  2006, pp. 11--18.

\bibitem{1323489}
N.~Abu-Ghazaleh, M.~J. Lewis, and M.~Govindaraju, ``Differential serialization
  for optimized soap performance,'' in \emph{Proceedings. 13th IEEE
  International Symposium on High performance Distributed Computing, 2004.},
  June 2004, pp. 55--64.

\bibitem{SOAP}
K.~Chiu, M.~Govindaraju, and R.~Bramley, ``Investigating the limits of soap
  performance for scientific computing,'' in \emph{Proceedings 11th IEEE
  International Symposium on High Performance Distributed Computing}, 2002, pp.
  246--254.

\bibitem{5695644}
C.~Wang, F.~Mueller, C.~Engelmann, and S.~L. Scott, ``Hybrid checkpointing for
  mpi jobs in hpc environments,'' in \emph{2010 IEEE 16th International
  Conference on Parallel and Distributed Systems}, Dec 2010, pp. 524--533.

\bibitem{1559961}
R.~Gioiosa, J.~C. Sancho, S.~Jiang, and F.~Petrini, ``Transparent, incremental
  checkpointing at kernel level: a foundation for fault tolerance for parallel
  computers,'' in \emph{Supercomputing, 2005. Proceedings of the ACM/IEEE SC
  2005 Conference}, Nov 2005, pp. 9--9.

\bibitem{5160999}
G.~Bronevetsky, D.~Marques, K.~Pingali, S.~McKee, and R.~Rugina,
  ``Compiler-enhanced incremental checkpointing for openmp applications,'' in
  \emph{2009 IEEE International Symposium on Parallel Distributed Processing},
  May 2009, pp. 1--12.

\bibitem{pbkl:95:lib}
J.~S. Plank, M.~Beck, G.~Kingsley, and K.~Li, ``{\bf Libckpt}: Transparent
  checkpointing under {Unix},'' in \emph{Usenix Winter Technical Conference},
  January 1995, pp. 213--223.

\bibitem{10.1007/978-3-642-24449-0_31}
K.~B. Ferreira, R.~Riesen, R.~Brighwell, P.~Bridges, and D.~Arnold,
  ``libhashckpt: Hash-based incremental checkpointing using gpu's,'' in
  \emph{Recent Advances in the Message Passing Interface}, Y.~Cotronis,
  A.~Danalis, D.~S. Nikolopoulos, and J.~Dongarra, Eds.\hskip 1em plus 0.5em
  minus 0.4em\relax Berlin, Heidelberg: Springer Berlin Heidelberg, 2011, pp.
  272--281.

\bibitem{kurtThesis}
K.~B. Ferreira, ``Keeping checkpointing viable for exascale systems,'' Ph.D.
  dissertation, The University of New Mexico, 2011.

\bibitem{libckptWeb}
``Libckpt home page,''
  \url{http://web.eecs.utk.edu/~plank/plank/www/libckpt.html}, accessed:
  2018-04-22.

\bibitem{bautista2011fti}
L.~Bautista-Gomez, S.~Tsuboi, D.~Komatitsch, F.~Cappello, N.~Maruyama, and
  S.~Matsuoka, ``{FTI: high performance Fault Tolerance Interface for hybrid
  systems},'' in \emph{{SC'11}}.

\bibitem{iniparser}
N.~Devillard, ``Iniparser 4,'' \url{https://github.com/ndevilla/iniparser},
  2017.

\bibitem{fti}
L.~Bautista-Gomez, ``Fti - fault tolerance interface,''
  \url{https://github.com/leobago/fti}, 2018.

\bibitem{FFF}
``Fti file format documentation,''
  \url{http://leobago.github.io/fti/ftiff.html}.

\bibitem{HARRAN2017}
\BIBentryALTinterwordspacing
M.~Harran, W.~Farrelly, and K.~Curran, ``A method for verifying integrity and
  authenticating digital media,'' \emph{Applied Computing and Informatics},
  2017. [Online]. Available:
  \url{http://www.sciencedirect.com/science/article/pii/S2210832717300753}
\BIBentrySTDinterwordspacing

\bibitem{zlib}
M.~Adler and J.-l. Gailly, ``Zlib data compression library,''
  \url{https://github.com/madler/zlib}, 1995-2017.

\bibitem{Nakassis1988}
\BIBentryALTinterwordspacing
A.~Nakassis, ``Fletcher's error detection algorithm: How to implement it
  efficiently and how toavoid the most common pitfalls,'' \emph{SIGCOMM Comput.
  Commun. Rev.}, vol.~18, no.~5, pp. 63--88, Oct. 1988. [Online]. Available:
  \url{http://doi.acm.org/10.1145/53644.53648}
\BIBentrySTDinterwordspacing

\bibitem{openssl}
``{OpenSSL} 1.0.2 manpages - md5,''
  \url{https://www.openssl.org/docs/man1.0.2/crypto/md5.html}, accessed:
  2018-04-17.

\bibitem{avalanche}
S.~Ramanujam and M.~Karuppiah, ``Designing an algorithm with high avalanche
  effect,'' \emph{IJCSNS International Journal of Computer Science and Network
  Security}, vol.~11, no.~1, pp. 106--111, 2011.

\bibitem{GUERON2012179}
\BIBentryALTinterwordspacing
S.~Gueron, ``Speeding up crc32c computations with intel crc32 instruction,''
  \emph{Information Processing Letters}, vol. 112, no.~5, pp. 179 -- 185, 2012.
  [Online]. Available:
  \url{http://www.sciencedirect.com/science/article/pii/S002001901100319X}
\BIBentrySTDinterwordspacing

\bibitem{zlibWeb}
``zlib home page,'' \url{https://zlib.net}, accessed: 2018-05-3.

\bibitem{mn4userguide}
``bsc.es, marenostrum4 user's guide,''
  \url{https://www.bsc.es/user-support/mn4.php#systemoverview}, accessed:
  2018-04-27.

\bibitem{mn4press}
``bsc.es, marenostrum iv (2017) system architecture,''
  \url{https://www.bsc.es/marenostrum/marenostrum/technical-information},
  accessed: 2018-04-27.

\bibitem{LULESH2:changes}
I.~Karlin, J.~Keasler, and R.~Neely, ``Lulesh 2.0 updates and changes,'' Tech.
  Rep. LLNL-TR-641973, August 2013.

\bibitem{LULESH:spec}
``{H}ydrodynamics {C}hallenge {P}roblem, {L}awrence {L}ivermore {N}ational
  {L}aboratory,'' Tech. Rep. LLNL-TR-490254.

\bibitem{MARKIDIS20101509}
\BIBentryALTinterwordspacing
S.~Markidis, G.~Lapenta, and Rizwan-uddin, ``Multi-scale simulations of plasma
  with ipic3d,'' \emph{Mathematics and Computers in Simulation}, vol.~80,
  no.~7, pp. 1509 -- 1519, 2010, multiscale modeling of moving interfaces in
  materials. [Online]. Available:
  \url{http://www.sciencedirect.com/science/article/pii/S0378475409002444}
\BIBentrySTDinterwordspacing

\bibitem{DEEPEST}
``Deep projects,'' \url{http://www.deep-projects.eu/}.

\bibitem{bib:boost}
``Boost - serialization,''
  \url{https://www.boost.org/doc/libs/1_67_0/libs/serialization/doc/index.html},
  accessed: 2018-05-17.

\bibitem{4536277}
W.~Yu, J.~S. Vetter, and H.~S. Oral, ``Performance characterization and
  optimization of parallel i/o on the cray xt,'' in \emph{2008 IEEE
  International Symposium on Parallel and Distributed Processing}, April 2008,
  pp. 1--11.

\bibitem{Shan2007UsingIT}
H.~Shan and J.~Shalf, ``Using ior to analyze the i / o performance for hpc
  platforms,'' 2007.

\bibitem{NvmeWopsss2018}
L.~B. Gomez, K.~Keller, and O.~Unsal, ``Performance study of non-volatile
  memories on a high-end supercomputer,'' in \emph{Workshop on Performance and
  Scalability of Storage Systems 2018 (WOPSSS'18), Frankfurt, Germany}, June
  2018.

\end{thebibliography}
